# Fourier-Tailored Light–Matter Coupling in van der Waals Heterostructures

*Dorte R. Danielsen[†][‡], Nolan Lassaline[†][‡], Sander J. Linde[†], Magnus V. Nielsen[†], Xavier Zambrana-Puyalto[†], Avishek Sarbajna[†], Duc Hieu Nguyen[†], Timothy J. Booth[†], Nicolas Stenger[§], Søren Raza*[†]*

[†]Department of Physics, Technical University of Denmark, 2800 Kongens Lyngby, Denmark

[§]Department of Electrical and Photonics Engineering, Technical University of Denmark, 2800 Kongens Lyngby, Denmark

ABSTRACT.

Dielectric structures can support low-absorption optical modes, which are attractive for engineering light–matter interactions with excitonic resonances in two-dimensional (2D) materials. However, the coupling strength is often limited by the electromagnetic field being confined inside the dielectric, reducing spatial overlap with the active excitonic material. Here, we demonstrate a scheme for enhanced light–matter coupling by embedding excitonic tungsten disulfide ($WS_2$) within dielectric hexagonal boron nitride (hBN), forming a van der Waals (vdW) heterostructure that optimizes the field overlap and alignment between excitons and optical waveguide modes. To tailor diffractive coupling between free-space light and the waveguide modes in the vdW heterostructure, we fabricate Fourier surfaces in the top hBN layer using thermal scanning-probe lithography and etching, producing sinusoidal topographic landscapes with nanometer precision. We observe the formation of exciton-polaritons with a Rabi splitting indicating that the system is at the onset of strong coupling. These results demonstrate the potential of Fourier-tailored vdW heterostructures for exploring advanced optoelectronic and quantum devices.

**INTRODUCTION**

The coherent exchange of energy between light confined in optical resonators and electronic transitions in materials can create hybrid polaritonic states that offer powerful opportunities to engineer optoelectronic materials. This regime of enhanced light–matter coupling has opened exciting opportunities across a range of fields,[1] such as creating advanced polaritonic devices[2,3] or fundamentally modifying chemical reactions.[4,5] These effects hold the potential to access and manipulate material properties at the quantum level.[6,7]

Achieving optimal light–matter coupling requires the transition dipole moment **μ** of the electronic resonance and the electric field **E** of the optical mode to be maximal, aligned, and overlap spatially and spectrally, as the coupling strength scales as $g \sim \mathbf{\mu} \cdot \mathbf{E}$. Transition metal dichalcogenides (TMDs) are particularly well-suited for this purpose, as they host excitons with strong in-plane dipole moments.[8] While this property has been combined with plasmonic resonances to reach the strong coupling limit,[9–12] high-index dielectric materials present an alternative platform to support near-absorption-free optical resonances.[13–15] However, in dielectric materials, the strongest electromagnetic fields are typically confined within the structure, limiting the spatial overlap of the optical modes with an external active material.

Van der Waals (vdW) heterostructures offer a solution to this challenge by allowing placement of the active TMD material within the dielectric material hexagonal boron nitride (hBN),[16–18] which is known for its extensive use as an encapsulant in vdW heterostructures.[19] hBN not only encapsulates 2D materials to protect against contamination and oxidation,[20–23] but also provides an atomically flat, defect-free interface[24–27] that preserves the rich structure of excitonic complexes in 2D TMDs.[28,29] Additionally, the wide transparency window (~6 eV bandgap)[30] and moderate refractive index of hBN make it an attractive platform for waveguiding applications in photonics.[31–33] The resulting confined waveguide modes feature strong in-plane electric fields that can overlap with TMD excitons. However, these modes cannot be directly excited from the far-field, necessitating efficient schemes to couple light into vdW heterostructures.

Optical Fourier surfaces provide a powerful and analytical approach to tailor diffractive coupling in nanophotonic devices.[34] Unlike conventional lithographic methods, which produce square-like gratings with limited control over spatial Fourier components[35,36], Fourier surfaces leverage the principle of wave superposition to enable precise and selective control over the Fourier components of a grating.[34,37] This precision is particularly advantageous in vdW heterostructures, where variations in the placement of 2D materials and the thicknesses of encapsulating hBN layers can impact device performance. By engineering the Fourier components to match the optical and structural requirements of a heterostructure, Fourier surfaces can ensure efficient coupling between far-field light and confined waveguide modes. This capability enables the excitation of hBN waveguide modes with strong in-plane fields and their coupling to excitons in TMDs, thereby enhancing light–matter interactions in these systems, while allowing for a simple far-field excitation and measurement scheme. Furthermore, Fourier surfaces can be exploited to introduce and engineer bandgaps in the optical dispersion, offering a platform for manipulating the photonic band structure in nanoscale optoelectronic devices.

Here, we pattern Fourier surfaces in hBN using thermal scanning probe lithography (tSPL),[38,39,40] allowing us to engineer the diffractive coupling between free space optical modes and confined waveguide modes in the vdW heterostructure. By encapsulating a monolayer of tungsten disulfide ($WS_2$) within two hBN flakes, we achieve enhanced light–matter coupling between waveguide modes supported by dielectric hBN flakes and excitons in the $WS_2$ monolayer. We observe a clear Rabi splitting, demonstrating exciton-polariton formation, with a magnitude indicating that the system is at the onset of strong coupling. This approach combines the precision of tSPL patterning, the excellent optical properties of hBN, and the strong excitonic response of TMDs. Our work demonstrates a versatile, integrated platform for engineering light–matter interactions in advanced vdW heterostructures.

**RESULTS AND DISCUSSION**

**Fourier Grating Couplers in hBN**

A dielectric slab of hBN supports both transverse electric (TE) and transverse magnetic (TM) waveguide modes, where the TE modes are characterized by only having an in-plane electric field component, and the TM modes have both in- and out-of-plane electric field components. We focus on the TE modes as these have the largest in-plane electric fields, suitable for coupling to in-plane excitons of TMDs.

The dispersion relations of the fundamental $TE^0$ and higher-order $TE^1$ modes in 150 nm thick hBN surrounded by air are depicted in Fig. 1a. The dispersion relations shown are based on the theory of a flat dielectric slab waveguide[41], assuming an isotropic constant refractive index of 2.1 in the hBN slab (corresponding to the in-plane refractive index of hBN in the visible range[42]). The $TE^0$ mode has the strongest optical confinement in the hBN waveguide, as evidenced by the larger in-plane momentum ($k_x$) and an electric field anti-node in the center of the waveguide with maximal electric field. In contrast, the $TE^1$ mode displays a node in the center of the waveguide with minimal electric field. For both bound modes, the in-plane momentum ($k_x$) exceeds that of free-space light, necessitating an efficient method for coupling light in and out of the waveguide structure through momentum matching.

By designing Fourier grating couplers based on sinusoidal superpositions, it is possible to precisely control diffracted wavefronts and realize efficient optical coupling into and out of hBN waveguides (Fig. 1b,c). The surface profile of the grating coupler provides additional momenta governed by its Fourier expansion.[34,43,44] As such, a grating surface composed of a sum of sinusoids, $f(x) = \sum_j A_j \cos(q_j x + \phi_j)$, will provide additional momenta corresponding to the wavevectors, $\mathbf{q}_j = (2\pi/\Lambda_j)\hat{\mathbf{x}}$, where $\Lambda_j$ are the spatial periods, allowing the incident electromagnetic waves to couple to waveguide modes in the hBN when $\mathbf{k}_{i,x} + \mathbf{q}_j = \mathbf{k}_{WG}$. Here $\mathbf{k}_{i,x}$ is the in-plane wavevector of the incident light and $\mathbf{k}_{WG}$ is the wavevector of the waveguide mode. When the Fourier surface grating consists of a single sinusoid $f(x) = A\cos(qx)$, points in $k$-space separated by $q$ will be coupled and free-space photons within the light cone can diffractively couple to the waveguide modes. This is illustrated by displacing the bands along the momentum axis by $q$ (see Fig. 1b,c). Since the waveguide mode dispersion is symmetric for counter-propagating modes, the shifted bands cross each other at $k_x = 0$. Furthermore, the periodic grating allows coupling between counter-propagating waveguide modes when their wavevectors are separated by $q$, leading to bandgaps in the mode dispersion. The bandgaps appear outside the light cone at $k_x = \pm q/2$. When the hBN slab also supports the $TE^1$ mode (Fig. 1c), band crossings of both the counter-propagating $TE^1$ and $TE^0$ modes appear inside the light cone, while bandgaps open in the $TE^1$ mode dispersion outside the light cone. Finally, the $TE^0$ and $TE^1$ modes couple when their mode dispersions are separated by $q$, leading to avoided crossings.

**Single sinusoid Fourier gratings**

We pattern single-sinusoid Fourier gratings with varying amplitudes and spatial frequencies in 130-nm-thick hBN (see Fig. 2a and Fig. S1-2). The grating topography is measured by atomic force microscopy (AFM) and the AFM line-profiles are fitted to the sinusoidal function $f(x) = A\cos(qx)$ (see Methods and Table S1). For the grating outlined by the white box in Fig. 2a, the average root-mean square error (RMSE) of the line fits is 1.95 nm, while the RMSE roughness of the surrounding unpatterned region is 0.9 nm, demonstrating nanometer precision of the patterned sinusoidal function (see Methods). After patterning, the hBN flake is transferred to a glass substrate and the optical response is characterized with angle-resolved reflectance spectroscopy (see Methods). This measurement technique captures spectrally dispersed Fourier-space patterns (energy $E$ vs. in-plane wavevector $k_x$) in reflectance mode, providing direct insight into the optical band structure of hBN flakes[40]. We observe distinct features in reflectance corresponding to replicas of the waveguide modes shifted into the optical window by the grating momentum (Fig. 2b). In addition, we observe increased reflectance around normal incidence ($k_x \approx 0$ μm$^{-1}$) due to light that is not diffracted into the first order by the grating. The presence of the waveguide modes in the observable optical range confirms the effectiveness of our grating structure in coupling free-space light to waveguide modes, as supported by reflectance simulations (Fig. 2c). The simulated reflectance spectra are obtained from finite-element simulations of an infinite grating structure with parameters based on the fitted experimental surface profile from AFM measurements (see Methods). The absence of additional bands and bandgaps in the optical measurement confirms the sinusoidally pure nature of the Fourier surface, without additional unwanted spatial frequencies.

The out-coupled light from the waveguide modes interferes with the spectrally broad background reflection of non-coupled waves from the hBN slab, giving rise to Fano-like lineshapes in reflectance[43] (see Fig. S3). The positions of the Fano-like lineshapes in experiments and simulations match, while the lineshapes are much broader in experiments than in simulations. Based on simulations with finite-sized gratings (see Fig. S3), we attribute the broadened lineshapes and reduced reflectance maximum in experiments to the finite size of the grating (15x15 μm). In general, lineshape broadening can also be a result of large grating amplitudes or imperfections in the grating. However, this grating has a very small sine amplitude (8.5 nm) and a small RMSE (1.95 nm), so we do not expect these factors to limit the resonance linewidth.

**Superimposed Fourier Gratings**

Photonic band-structure engineering, achieved through surface design and tSPL patterning, provides a robust means of designing custom optical responses in vdW heterostructures. A single sinusoidal grating structure couples free-space light to waveguide modes and vice versa, generating a replica of the waveguide band structure that produces crossing bands within the light cone, while also producing bandgaps in the dispersion outside the light cone (Fig. 1a). By including a second sinusoid in the surface pattern, it is possible to fold the bandgaps into the light cone (Fig. 3a). In this example, we patterned a Fourier surface grating with $f(x) = A_1 \cos(q_1 x) + A_2 \cos\left(q_2 x - \frac{\pi}{2}\right)$, where $q_1 = 2\pi/\Lambda$ and $q_2 = 2q_1$ (Fig. 3b). Here, the second cosine will create bandgaps in the mode dispersion at $k_x = \pm q_2/2 = q_1$, and the first cosine will fold the bandgaps to the center of the light cone at $k_x = 0$, as illustrated in Fig. 3a.

Our experimental results show reflectance peaks that correspond to the dispersions for both the fundamental ($TE^0$) and higher-order ($TE^1$) waveguide modes, alongside observable bandgaps due to the double-sinusoid surface profile (Fig. 3c). The measured data in Fig. 3c are obtained from the vdW heterostructure shown in Fig. 4a from a region where only hBN is present. The experimental results align closely with reflectance simulations (Fig. 3d), validating the observed effects as arising from Fourier-engineered diffractive properties. The simulated electric field profiles in Fig. 3e show that the lower-energy bands are indeed associated with the $TE^0$ mode as there are no nodes in the field profiles of the upper (UB) and lower (LB) branches at the bandgap. The higher-order bands exhibit a node in the field profiles and are thus associated with the $TE^1$ mode (Fig. 3e).

**Excitonic Heterostructures**

To explore light–matter interactions within vdW heterostructures, we encapsulate a monolayer of $WS_2$ between two hBN flakes, where a Fourier grating is patterned in the top hBN flake (Fig. 4a). This configuration combines the optical waveguide modes in hBN with the excitons in $WS_2$, creating a platform that enables tailored light–matter coupling. Furthermore, the excitonic material is protected from the external environment both during fabrication and subsequent optical measurements. In this device the $WS_2$ is present under only part of the grating, which allows us to measure the dispersion with and without $WS_2$ from the same grating (Fig. 4a). The two-component sinusoid shown in Fig. 3 is used as the Fourier grating (Fig. 4b, Table S1).

Optical measurements are performed using angle-resolved reflectance spectroscopy, where the specific regions of the grating with and without $WS_2$ are selected by moving the position of a real-space aperture with respect to the sample (see Methods). In this way we can directly observe the impact of the excitonic layer on the measured optical response from the vdW heterostructure with the Fourier grating. For direct comparison, in Fig. 4c the spectrum on the left ($k_x<0$) is from the grating region with $WS_2$, and the spectrum on the right ($k_x>0$) is from the area without $WS_2$. Even though the sign of $k_x$ differs in the two cases, these spectra are comparable as they correspond to the same propagation direction with respect to the edge of the grating (see Fig. S4). The experimental reflectance results agree well with the simulations (Fig. 4d), where the fitted parameters from AFM measurements are used as input for the simulation geometry. When $WS_2$ is present, both experiments (Fig. 4c) and simulations (Fig. 4d) exhibit a clear reflectance dip, independent of $k_x$, corresponding to the A exciton resonance of $WS_2$[8,45]. As the reflectance peak corresponding to the $TE^0$ waveguide mode approaches the exciton resonance, an anti-crossing is visible both in experiments and simulations. For comparison, there is no visible anti-crossing when $WS_2$ is absent (Fig. 4c,d). Furthermore, with $WS_2$ in the heterostructure, the reflectance features corresponding to the $TE^0$ mode dispersion are fainter above the exciton energy compared to the $TE^0$ mode without $WS_2$. This is most clear in the simulated reflectance spectra but is also evident from the experimental spectra. On the other hand, the $TE^1$ mode dispersion is largely the same both with and without $WS_2$ in the heterostructure.

To gain further insights on the interactions between WS$_2$ excitons and the waveguide modes, we plot the electric field profiles of the TE$^0$ and TE$^1$ modes of the vdW heterostructure (Fig. 4e,f) corresponding to energies and in-plane *k*-vectors indicated by the white arrows in Fig. 4d. The field profiles show that the electric field of the TE$^0$ mode is maximal at the position of the WS$_2$ layer (Fig. 4e), while the electric field of the TE$^1$ mode is almost minimal at the WS$_2$ layer due to the node of the TE$^1$ mode. Placing the WS$_2$ layer at the position in the vdW heterostructure where the TE$^0$ electric field is maximal allows us to maximize the coupling strength between WS$_2$ excitons and the TE$^0$ mode, while minimizing coupling to the TE$^1$ mode. This can also explain the modified reflectance of the TE$^0$ mode above the exciton energy, as the presence of WS$_2$ leads to absorption losses. Since the electric field of the TE$^1$ mode is minimal at WS$_2$ layer, the TE$^1$ mode is largely unaffected by the presence of WS$_2$.

**Onset of Strong Coupling**

The avoided crossing observed in the presence of WS$_2$ in Fig. 4 is a signature of strong light–matter coupling between the waveguide mode and the excitons in WS$_2$. To quantify the observed light–matter coupling, we extract values for the Rabi splitting $E_{Rabi}$ and the coupling strength *g*. To do this, we first determine the resonance energies and the linewidths of the uncoupled exciton and waveguide mode. We use the waveguide mode dispersion from the region without WS$_2$ to extract the uncoupled waveguide mode. The uncoupled exciton resonance is determined from reflectance spectra obtained at large angles ($k_x/k_0$ between -0.79 and -0.47, with $k_0 \approx 10$ μm$^{-1}$), where the exciton is far away from the waveguide mode. In both cases the Fano lineshapes that appear in reflectance are due to the interference between a spectrally broad background reflectance (also observable in Fig. 4c,d) and the sharp waveguide or exciton resonances. The lineshapes are fitted to two interfering Lorentzian oscillators (see Methods) to extract the resonance energies and linewidths of the uncoupled systems (see Fig. 5a,b). The resulting exciton resonance energy is 1984.3 ± 0.6 meV, which is consistent with previous reports of the A exciton resonance in WS$_2$[45]. The extracted half-width-at-half-maximum (HWHM) of the exciton and hBN waveguide mode are $\gamma_{Ex}$ = 15.1±0.6 meV and $\gamma_{Cav}$ = 33.8±0.2 meV, respectively. Similarly, we determine the energies of the upper and lower branches at the anti-crossing by fitting the reflectance spectra in Fig. 5c to a sum of three interfering Lorentzian oscillators (see Methods). The resulting energy dispersions of the upper and lower branches as well as the uncoupled exciton and hBN waveguide mode are shown in Fig. 5d. The Rabi splitting is estimated as the energy difference between the upper and lower branches of the coupled system at the point where the uncoupled exciton and hBN waveguide mode intersect, yielding $E_{Rabi}$ = 40±9 meV (see Methods). Based on the coupled oscillator model, the coupling strength can be calculated from the Rabi splitting and the uncoupled linewidths, yielding $g$ = 22±4 meV (see Methods). Similar results for $E_{Rabi}$ and $g$ were obtained with a different approach where the uncoupled waveguide mode is estimated based on energy conservation considerations (see Fig. S5 and Table S2). A commonly used criterion for classifying strong coupling is that both $E_{Rabi} > \gamma_{Cav} + \gamma_{Ex}$ and $2g > |\gamma_{Cav} - \gamma_{Ex}|$ should be fulfilled[46,47], corresponding to a minimum of one Rabi oscillation[48]. In our system, the former condition is not satisfied while the latter is satisfied (see Table S2). As such, we assess that the coupled exciton–waveguide system is at the onset of the strong coupling regime, since there is a visible anti-crossing in reflectance, but the extracted Rabi splitting does not quite exceed the overall losses of the system.

## CONCLUSIONS

In this study, we engineered light–matter interactions by integrating Fourier surfaces within vdW heterostructures containing hBN and $WS_2$. This approach enables diffractive coupling between free-space optical modes and bound waveguide modes, as well as coupling between counter-propagating modes to produce bandgaps in the optical dispersion. By embedding a monolayer of $WS_2$ inside the vdW heterostructure, it is possible to protect the excitonic 2D material from the external environment. Simultaneously, this positions the excitonic material at the location in the hBN waveguide where the electric field of the $TE^0$ mode is maximal. This configuration enables enhanced light–matter interactions, as evidenced by the observed anti-crossing in reflectance measurements due to coupling between excitonic states in $WS_2$ and waveguide modes in hBN. Based on the extracted values for the Rabi splitting, coupling strength, and linewidths of the uncoupled systems we assess that the coupled exciton–waveguide mode is at the onset of the strong coupling regime. The light–matter interactions in our system could be further enhanced, for example, by increasing the number of $WS_2$ layers[10,11,46,49] or tuning the bandgap in the $TE^0$ mode dispersion such that a band edge coincides with the exciton resonance, thereby exploiting slow light[50]. Combining Fourier surfaces with vdW heterostructures is a compelling platform for fundamental studies of light–matter interactions and for the development of advanced optoelectronic devices.

## METHODS

### Mechanical Exfoliation

Flakes of hBN from HQ Graphene were exfoliated onto 90 nm $SiO_2$/Si substrates using tape from Nitto (1007R Silicone-Free Blue Adhesive Plastic Film – Medium tack, 5.9'' from Ultron Systems and EPL BT-150E-KL from Nitto). The flakes were visually inspected under an optical microscope, and their thickness was determined through atomic force microscopy (AFM) and optical contrast. $WS_2$ from HQ Graphene was similarly exfoliated onto 90 nm $SiO_2$/Si, using scotch tape (3M Scotch® Magic$^{TM}$ tape).

### tSPL Patterning

Thermal scanning-probe lithography (tSPL) was employed to create the Fourier surface patterns on the hBN flakes. Before spin coating, an oxygen plasma cleaning step was used to improve resist adhesion. Oxygen plasma cleaning was performed in a capacitively coupled plasma system (MiniLab 026 Soft Etching system from Moorfield). The $O_2$ flow rate was set to 15 sccm, and the power was set 30 W, which resulted in a DC bias of approximately 228 V, and pressure of approximately 0.015 mbar. The stabilization time was 2 s and the cleaning time was 120 s. A resist of 9 wt% polyphthalaldehyde (PPA) in anisole was spin-coated onto the surface of the $SiO_2$/Si chips containing hBN flakes at a spin speed of 6000 r.p.m. and 2000 r.p.m./s acceleration for 40 s to achieve a uniform layer (approximately 150 nm thick). After spin coating, the sample was baked on a hot plate for two minutes at 110 °C. The resist was patterned by tSPL using the Nanofrazor system (Heidelberg instruments) to selectively remove the resist in a sinusoidal pattern[51]. The cantilever temperature is set to 1100 °C, and the set height of the cantilever is 225nm. The forces between the tip and the sample are controlled via the applied bias using a Kalman feedback loop.

**Etching**

The grayscale surface pattern in the PPA resist was transferred to underlying hBN flakes through reactive ion etching with $SF_6$ plasma. The etching was performed in a MiniLab 026 Soft Etching system from Moorfield (the same system that was used for oxygen plasma cleaning). The parameter setpoints are: $SF_6$ flow rate: 15 sccm and DC bias: 40 V, which results in a power of approximately 24 W and pressure of approximately 5 mbar). The etch time varies according to the required etch depth. In each run, the stabilization time is around 8 s. Following the etching process, the sample was cleaned with oxygen plasma (with the same process parameters as the oxygen cleaning process used before spin coating) to remove any remaining resist and contaminants, resulting in well-defined grating structures in the hBN for optical coupling.

**vdW Heterostructure Assembly**

VdW heterostructure assembly was performed using a standard dry-transfer technique[52,53] with a polycarbonate (PC)/polydimethylsiloxane (PDMS) stamp and a transfer system from hq graphene. The same technique was used to transfer flakes and vdW heterostructures to glass substrates. The overall fabrication process is illustrated in Fig. S1.

**Angle-resolved reflectance spectroscopy**

Angle-resolved reflectance spectroscopy was used to analyze the optical properties of the patterned waveguide. A $k$-space imaging setup consisting of three free-space lenses was integrated with an optical microscope (Nikon Eclipse LV100ND) and spectrograph (Andor Kymera 328i) to capture energy vs. $k_x$ reflectance maps (see illustration of the optical setup in Fig. S6). The optical setup is similar to that used in the literature[34,54].

A white-light source (Thorlabs OSL2) is used. The incident light is focused onto the sample with a 100x objective (Nikon, NA = 0.90) and the reflected light is collected through the same objective and passed through the tube lens of the microscope. A linear polarizing filter is integrated in the optical microscope to select reflected light with a given polarization angle, and thereby selectively measure the reflectance from TE modes. An adjustable iris is placed in the real-space image plane after the tube lens and the polarizing filter to ensure that the collected light only comes from the region of interest at the grating. The reflected light is passed through an additional set of lenses to project the back focal plane (i.e. the Fourier transform of the real-space sample image) onto the entrance slit of the spectrograph. In the Fourier-transformed image, the entrance slit limits the range of recorded $k$-vectors of the reflected light. The slit is centered around $k_{\text{x-pixel}} \approx 0$, which means that each $y$-pixel along the slit corresponds to a specific $k$-vector $(k_{\text{x-pixel}}, k_{\text{y-pixel}}) \approx (0, k_x)$, as illustrated in Fig S7. That is, at each $y$-pixel, a spectrum of light is recorded with a specific in-plane $k$-vector (i.e., light reflected at a specific angle) from all points of the sample, enclosed by the iris. The range of collected angles is dictated by the NA of the objective, with the maximal incident and reflected angle $\theta$ given by: $NA = n \sin(\theta)$, thus the generated reflectance maps span angles ranging from -65° to 65°. The reflectance, $R$, is obtained by normalizing the reflected signal from Fourier gratings, $I_{\text{hBN}}$, to the reflected signal from a mirror, $I_{\text{mirror}}$, and subtracting dark counts, $I_{\text{dark}}$, where no light is incident on the spectrometer:

$$R = \frac{I_{\text{hBN}} - I_{\text{dark}}}{I_{\text{mirror}} - I_{\text{dark}}} \ .$$

**AFM Topography Characterization**

The topographical features of the patterned hBN waveguides were characterized using AFM (Dimension Icon-PT AFM from Bruker AXS) in tapping mode. Before tSPL patterning the flake thicknesses were measured with AFM. After tSPL patterning and RIE etching, the surface topography of the Fourier gratings and the final flake thickness are measured with AFM. The measured surface topography data is fitted to the target sinusoidal functions to extract the amplitudes and spatial frequencies of the patterned Fourier surfaces. The *FunFit* software package[55] was used to determine the surface roughness and the root-mean-square error (RMSE) between the fitted functions and experimental topography. In this work, we analyze the surface topographies by examining the statistical average of line-by-line fits, in contrast to previous reports in the literature that use full 2D fits on the entire grating surface. This approach is taken to exclude the impact of cracks and bubbles in the heterostructures, thereby better reflecting the local grating structure in a given measurement on a selected area. We have cross-checked the RMSE values that we get with a full 2D fit, and we find that the quantities are comparable to those extracted from the average of the 1D fits, supporting the validity of our approach.

**Simulations**

Electromagnetic finite-element-method simulations were performed using COMSOL Multiphysics version 6.2 (wave optics module) to obtain simulated reflectance spectra and electric field profiles from the hBN Fourier grating couplers, both with and without $WS_2$ present. Due to the symmetry of the gratings, they are modelled with a 2D simulation geometry, where fitted parameters from the AFM measurements of the Fourier surface profile are used as input for the simulated surface profile in hBN. The hBN or hBN-$WS_2$-hBN heterostructures are on a glass substrate (700 nm thick with periodic port at the bottom for the outgoing waves), while the region above the hBN is filled with air. The grating is modelled as infinitely large using Floquet periodic boundary conditions. The anisotropic refractive index of hBN[42] and the refractive index of monolayer $WS_2$[8] are taken from literature. The refractive index of the glass is taken from an online database (https://refractiveindex.info/download/data/2014/BOROFLOAT%2033_Refractive%20Index. pdf.). The refractive index of air is set to 1. The system is excited by plane waves with angles of incidence between -65° and 65°, using a periodic port placed 1000 nm above the hBN surface, and the reflectance is determined at the same port. Electric field profiles as well as reflectance maps both from simulations and experiments are plotted with color maps from literature[56].

**Analysis of coupling strength using the coupled oscillator model**

To extract the energies and linewidths of the uncoupled exciton, we used reflectance spectra from the hBN-$WS_2$-hBN heterostructure at large angles ($k_x/k_0$ between -0.79 and -0.47). For the uncoupled waveguide mode, we used reflectance spectra from the grating region without WS$_2$. The reflectance $R$ exhibit Fano-like lineshapes at resonance, and were fitted to two interfering Lorentzian oscillators:

$$R(E) = \left| \frac{r_D \gamma_D}{i(E - E_D) + \gamma_D} e^{i\phi} + \frac{r_R \gamma_R}{i(E - E_R) + \gamma_R} \right|^2,$$

where $\phi$ is the relative phase between the two Lorentzians, $r_{D,R}$ is the amplitude, $E_{D,R}$ is the resonance energy, and $\gamma_{D,R}$ is the half-width-at-half-maximum ($2\gamma_{D,R}$ = FWHM) of the Lorentzian contributions from the spectrally broad background reflectance (D) and the resonant mode (R) (see Fig. 5a,b). The uncertainties of the fitted values are estimated as the 95% confidence intervals. The reported values for the uncoupled exciton energy $E_{Ex}$, and the half linewidths of the exciton (hBN waveguide mode) $\gamma_{Ex(Cav)}$ are the weighted averages of fitted values from several spectra, using the uncertainties as weights.

Similarly, the energies of the upper and lower branches near the anti-crossing (Fig. 5c) were determined by fitting to three interfering Lorentzian oscillators:

$$R(E) = \left| \frac{r_D \gamma_D}{i(E - E_D) + \gamma_D} + \frac{r_{R1} \gamma_{R1}}{i(E - E_{R1}) + \gamma_{R1}} e^{i\phi_1} + \frac{r_{R2} \gamma_{R2}}{i(E - E_{R2}) + \gamma_{R2}} e^{i\phi_2} \right|^2.$$

Assuming that the energies of the uncoupled waveguide mode $E_{Cav}$ can be approximated by the mode dispersion from the region with only hBN in the heterostructure, we found the point of zero detuning $\delta = E_{Cav} - E_{Ex}$ from the intersection between the exciton and hBN waveguide mode dispersion. The Rabi splitting $E_{Rabi}$ was then extracted as the energy difference between the upper and lower branches at the point of zero detuning. Based on the coupled oscillator model, the coupling strength $g$ was determined from:

$$E_{Rabi} = \sqrt{4g^2 - (\gamma_{Cav} - \gamma_{Ex})^2}.$$

The energies of the uncoupled waveguide mode were also extracted by invoking energy conservation in the region where the upper and lower branches overlap: $E_{Cav} = E_+ + E_- - E_{Ex}$. The energies of the upper and lower branches $E_\pm$ were then fitted to the coupled oscillator model to extract $g$ and $E_{Rabi}$. This approach led to similar results for $g$ and $E_{Rabi}$ (see Fig. S5 and Table S2).

# AUTHOR INFORMATION

**Corresponding Author**

*Email: sraz@dtu.dk

**Author Contributions.**

‡ These authors contributed equally (Dorte R. Danielsen and Nolan Lassaline)

N. L., M. V. N., S. J. L., D. R. D., and A. S. prepared samples for patterning. N. L. performed tSPL and etching. D. R. D., M. V. N. and S. J. L. performed dry-transfer. D. R. D., M. V. N., S. J. L., and A. S. carried out AFM and reflectance measurements. D. R. D. analyzed the reflectance spectra and carried out the coupling strength analysis. D. R. D. carried out AFM data analysis with input from N. L., M. V. N., and S. J. L. X. Z. P. built the optical setup for angle-resolved reflectance measurements with input from N. L. D. R. D., M. V. N., and S. R. carried out COMSOL simulations. D. H. N. provided exfoliated hBN. T.B. and N. S. co-supervised the project. S. R. supervised the project, and N. L. and S. R. conceived the idea. D. R. D., N. L. and S. R. wrote the manuscript, and all authors revised the manuscript.

**Acknowledgements**

We thank Peter Bøggild and Frederik Schröder for fruitful discussions, and Camilla Holm Sørensen for her assistance in optimizing the etching process.

**Funding**

N. L. acknowledges funding from the Swiss National Science Foundation (*Postdoc Mobility* P500PT_211105) and the Villum Foundation (*Villum Experiment* 50355). T. J. B. acknowledges support from the Novo Nordisk Foundation (BIOMAG NNF21OC0066526). D. R. D., A. S., S. R, T. J. B., and N. S. acknowledge funding from the Independent Research Funding Denmark (1032-00496B). X. Z.-P. and S. R. acknowledge funding from the Villum Foundation (VIL50376). N. S. thanks the Novo Nordisk Foundation NERD Programme (project QuDec NNF23OC0082957).

**Figures**

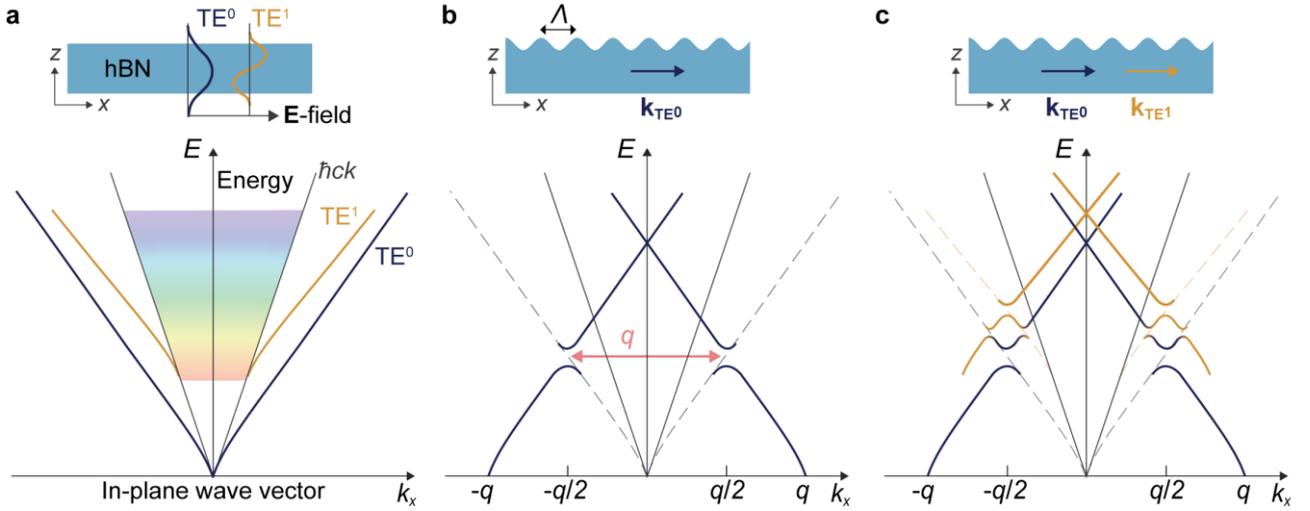

**Figure 1. Coupling to waveguide modes in a hBN slab.** (a) Top: Schematic of the $TE^0$ (blue line) and $TE^1$ (yellow line) modes in a flat dielectric slab waveguide of hBN. The TE modes only have electric field components in the in-plane direction (along the *y*-axis). Bottom: $TE^0$ and $TE^1$ mode dispersion curves in a flat hBN slab (here computed for a slab with isotropic refractive index $n = 2.1$ surrounded by air, $n = 1$). Propagating free space photons are bound by the light line ($E = \hbar ck$). The rainbow-colored trapezoid illustrates the accessible spectral region in reflectance measurements. (b) A sinusoidal modulation in the waveguide thickness acts as a diffraction grating that provides additional in-plane momentum $q = 2\pi/\Lambda$ to free-space optical modes, such that they can couple to waveguide modes with propagation vector $\mathbf{k}_{TE0}$ ($TE^0$, blue). The sinusoidal grating generates copies of the mode dispersion that are shifted by *q*, leading to band crossings within the light lines. Furthermore, counter-propagating modes couple when $q = 2\,|\mathbf{k}_{TE0}|$, thereby opening bandgaps outside the light lines. (c) Thicker waveguides support higher-order modes ($TE^1$, yellow), which also exhibit band crossings and bandgap openings, because of the single sinusoid grating. In addition, coupling occurs between counter-propagating $TE^0$ and $TE^1$ modes whose *k*-vectors are separated by *q*.

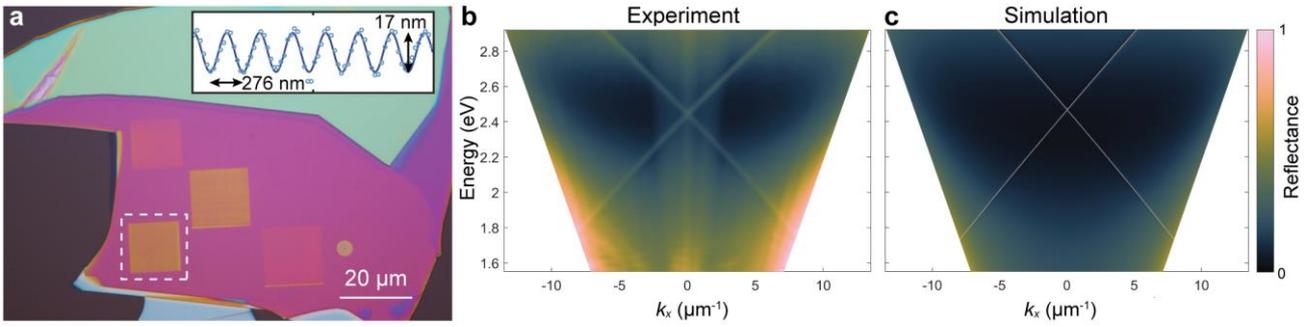

**Figure 2. hBN flake with a sinusoidal surface pattern.** (a) Bright-field microscopy image of a hBN flake with patterned sinusoidal gratings (squares, top view) on a $SiO_2$/Si substrate. Inset: Topography (cross-sectional view) of a grating coupler from the flake (dashed box), measured by AFM. The blue line is a fitted sinusoidal function showing a good agreement between the target profile and the fabricated structure. (b) Measured angle-resolved reflectance (energy vs. in-plane wave vector $k_x$) from the grating in (a), which has been transferred to a glass substrate. Two crossing streaks are observed, which correspond to diffractive coupling of free-space modes to $TE^0$ waveguide modes in the hBN flake. (c) Simulation of the experiment in (b) (see Methods).

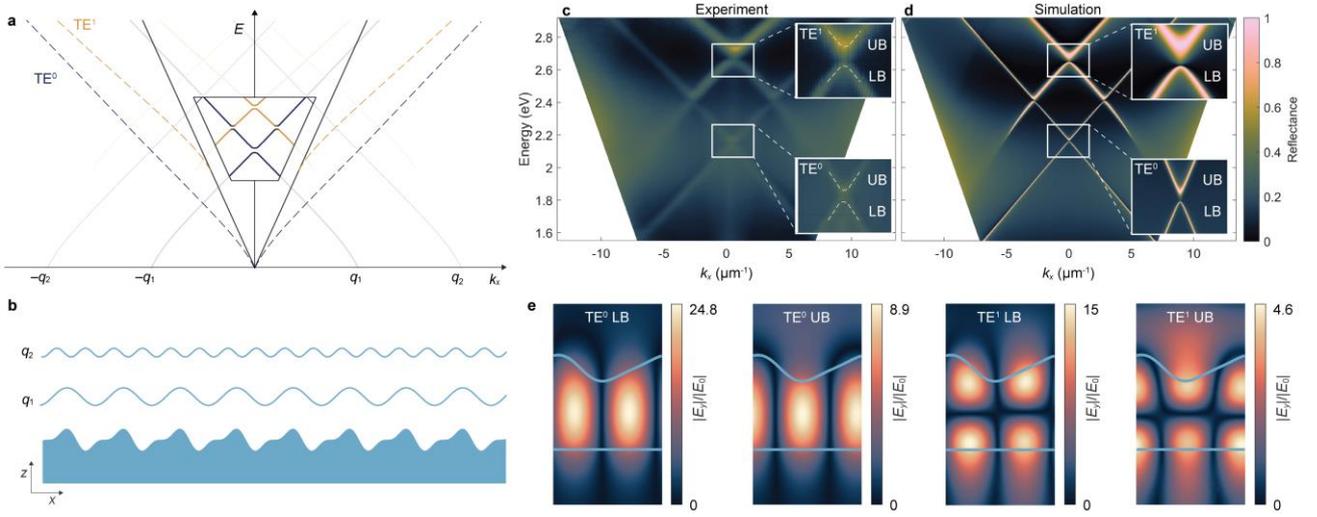

**Figure 3. Optical band-structure engineering in patterned hBN waveguides.** (a) Band-structure diagram for a hBN waveguide that supports $TE^0$ and $TE^1$ modes, with the surface profile in (b), composed of two superimposed sinusoids. The long-wavelength component of the surface profile generates a copy of the optical dispersion in the observable window, and the short-wavelength component opens bandgaps in the optical dispersion that are shifted into the observable window. (c) Measured angle-resolved reflectance from hBN with the surface profile in (b), where the theoretical predictions shown in (a) are observed. Insets: Zoomed-in plots on the band edges with UB (LB) indicating the upper (lower) band edge. The white dashed lines are guides to the eye. (d) Simulation of the experiment performed in (c). (e) Simulations of the electric-field distributions for the $TE^0$ and $TE^1$ modes at the upper and lower band edges in the dispersion diagram at $k_x = 0$.

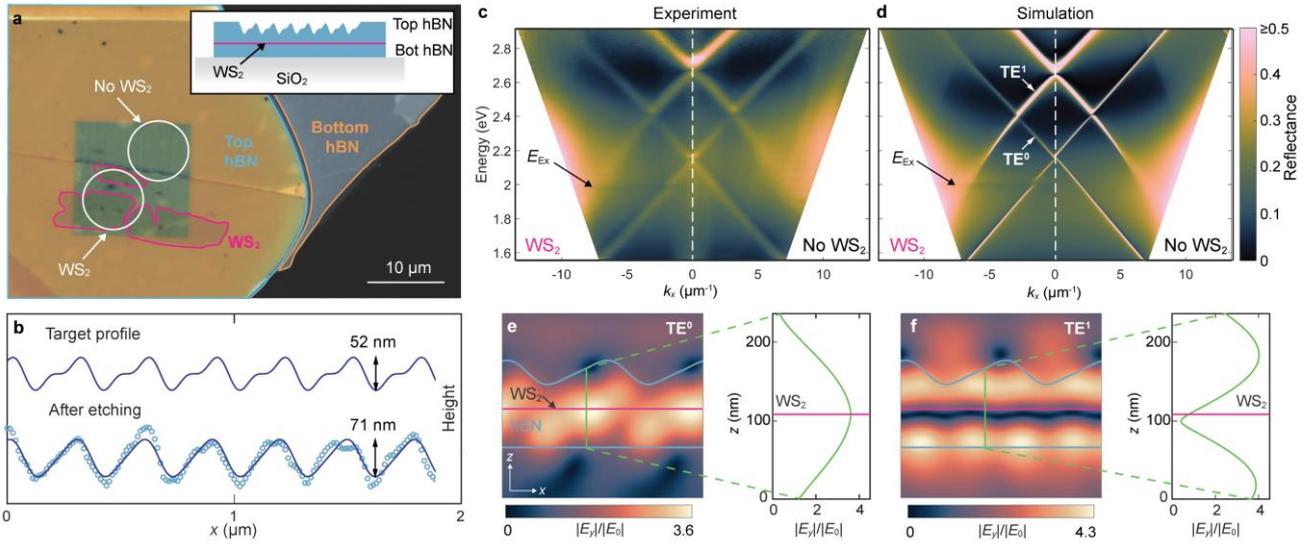

**Figure 4. Light–matter coupling in patterned vdW heterostructures.** (a) Bright-field microscopy image of a vdW heterostructure with monolayer $WS_2$ (pink) encapsulated in hBN. The bottom hBN (orange) is 110 nm thick, and the average thickness of the patterned top hBN (blue) is also 110 nm. $WS_2$ partially fills the grating region, allowing for direct measurement of the optical mode dispersion with and without the excitonic material (white circles). The inset shows a schematic cross-section of the patterned vdW heterostructure. (b) Target surface profile and AFM measured surface topography (blue dots) for the grating coupler in (a). The fitted blue line is used as the surface profile in simulations. (c) Measured angle-resolved reflectance from the vdW heterostructure in (a), where the left (right) half of the figure is measured on the grating where $WS_2$ is present (not present). The exciton resonance is visible as a horizontal reflectance dip indicated with the black arrow. (d) Simulation of the experiment performed in (c), where the $TE^0$ and $TE^1$ modes are indicated with white arrows. (e) In-plane electric-field distributions in the grating structure with $WS_2$ (pink) for the $TE^0$ mode from reflectance simulations indicated by the white arrow in (d). The inset to the right is the electric field along the green line cut, showing that the vertical position of $WS_2$ aligns with the field maxima of the $TE^0$ mode. (f) As in (e), but for the $TE^1$ mode. Here, the vertical position of $WS_2$ aligns with the field minima of the $TE^1$ mode.

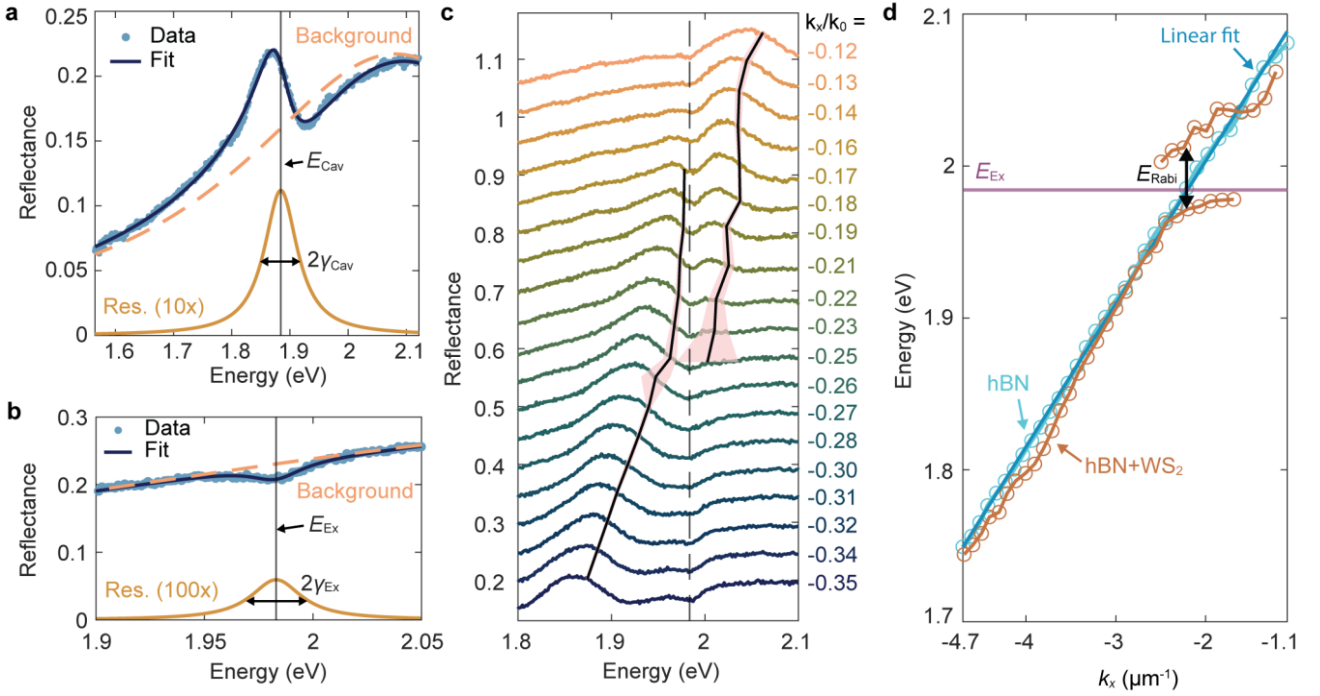

**Figure 5. Analysis of light–matter coupling strength.** (a) Reflectance spectrum (blue dots) from the hBN waveguide without WS$_2$ from Fig. 4 (obtained at $k_x/k_0 = -0.35$), fitted to two interfering Lorentzians (dark blue line) to consider Lorentzian contributions from a resonant waveguide mode (yellow line) and a broad background (orange dashed line). The resonant contribution is multiplied by a factor of 10 for clarity. (b) Reflectance spectrum near the exciton energy from the hBN waveguide with WS$_2$, obtained at large angles ($k_x/k_0 = -0.55$), where the exciton resonance is far from the waveguide mode. The spectrum is also fitted to two interfering Lorentzians to extract the exciton energy and linewidth. The resonant contribution is multiplied by a factor of 100 for clarity. (c) Reflectance data around the anti-crossing for a series of $k_x/k_0$ values. The resonance energies of the upper and lower branches are extracted by fitting the reflectance in this spectral region to three interfering Lorentzians (corresponding to the two branches and the background). The fitted energies are marked by the black lines and the 95% confidence interval of the fits are marked by the red shadow. (d) Mode dispersion for the coupled system with hBN and WS$_2$ (orange) from the resonance energies indicated in (c), together with the mode dispersion of the uncoupled hBN waveguide mode (blue) and the exciton resonance (purple). The dark blue line is a linear fit to the uncoupled hBN mode dispersion, indicating that the dispersion is linear in this region. The Rabi splitting is the energy difference between the upper and lower branches of the coupled system at the intersection between the exciton and the uncoupled hBN waveguide mode. The Rabi splitting is extracted from the data points of the upper and lower branches whose $k_x$-values are closest to that of the intersection.

# Supplementary Material for

# Fourier-Tailored Light–Matter Coupling in van der Waals Heterostructures

*Dorte R. Danielsen[†][‡], Nolan Lassaline[†][‡], Sander J. Linde[†], Magnus V. Nielsen[†], Xavier Zambrana-Puyalto[†], Avishek Sarbajna[†], Duc Hieu Nguyen[†], Timothy J. Booth[†], Nicolas Stenger[§], Søren Raza\*[†]*

‡These authors contributed equally (Dorte R. Danielsen and Nolan Lassaline)

∗Corresponding author: Søren Raza, E-mail: sraz@dtu.dk

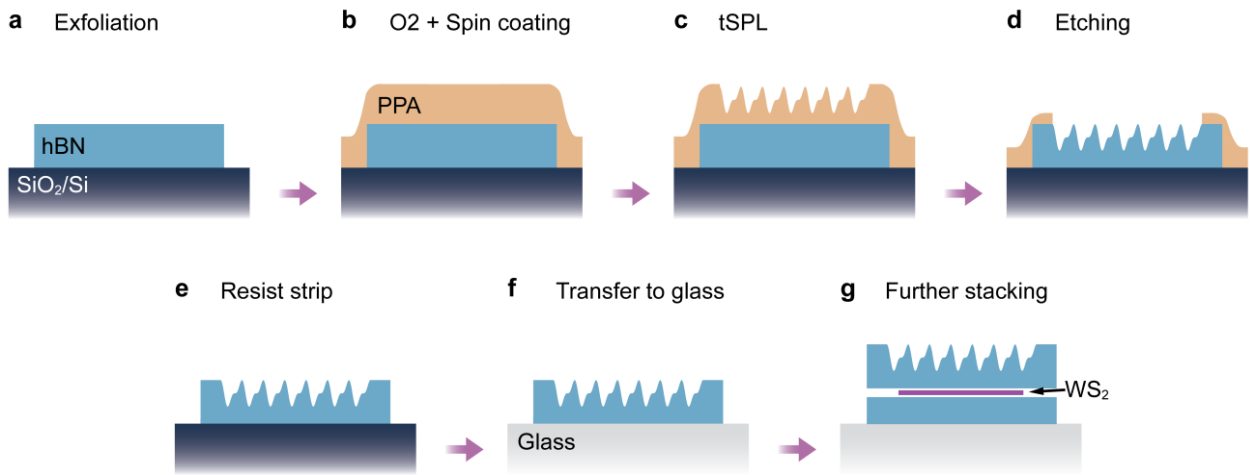

**Figure S1:** Schematic illustration of the fabrication process. (a) hBN is mechanically exfoliated onto $SiO_2$/Si. (b) PPA is spin coated onto the hBN flakes. Before spin coating, the sample is cleaned with oxygen plasma to remove residues and improve PPA adhesion. (c) Fourier surfaces are patterned in the PPA resist with tSPL. (d) The resist pattern is transferred to hBN using $SF_6$ etching. (e) The resist is removed by sublimation on a hot plate, and residues are removed with oxygen plasma. (f) The hBN is transferred to glass to improve the reflectance contrast. (g) The patterned hBN is used as the top-layer in a vdW heterostructure.

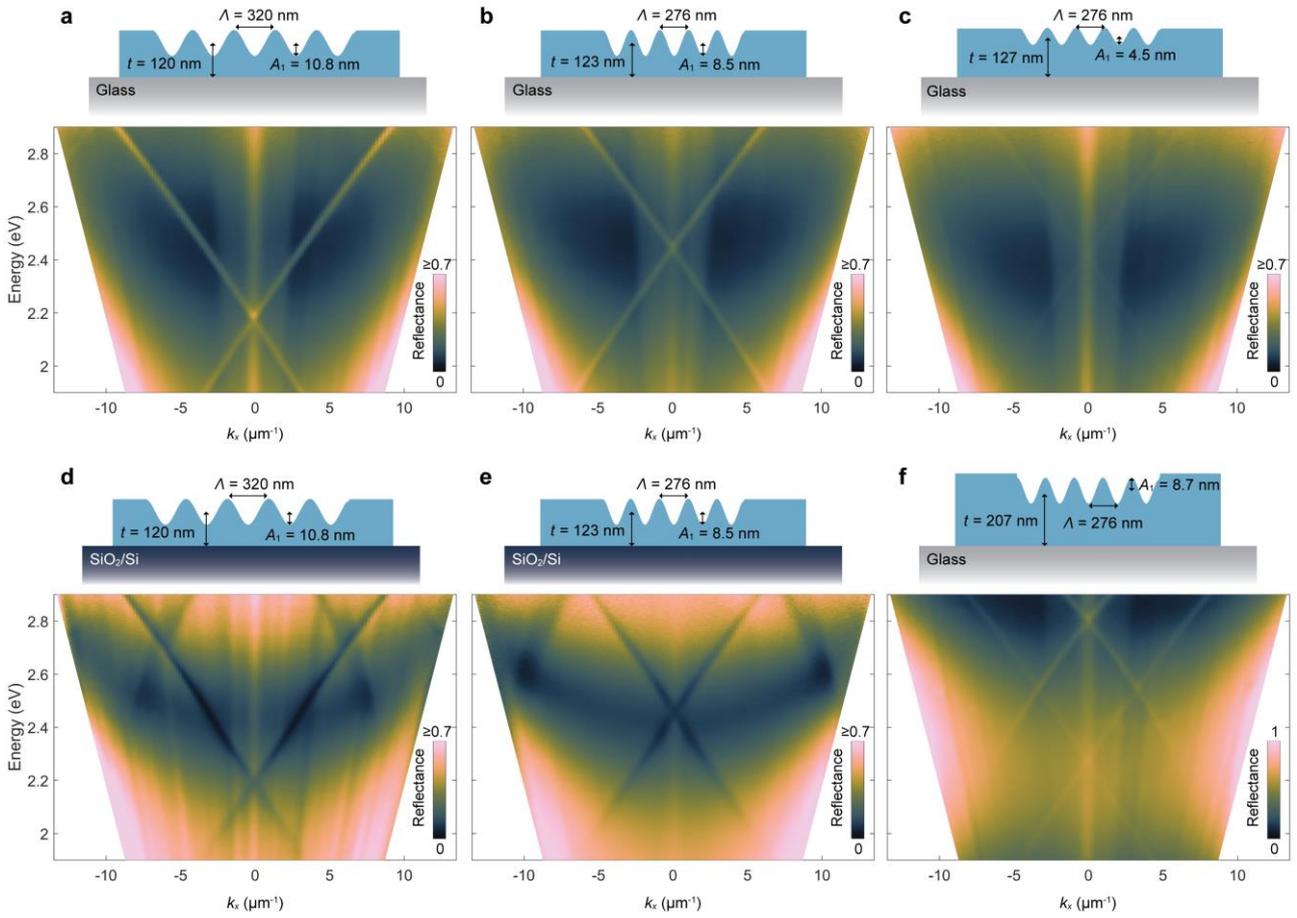

**Figure S2:** Angle-resolved reflectance from single sinusoids in hBN with various spatial periods $\Lambda$, amplitudes $A_1$, thicknesses $t$, and substrates, as indicated on the schematic illustrations in the top panels. (a-b) The hBN Fourier gratings have similar values for $A_1$ and $t$, but different $\Lambda$. Comparing (a-b) it is seen that increasing $\Lambda$ shifts the band crossing to lower energies. (c) The Fourier surface is similar to the one in (b), except $A_1$ is smaller. As a result, the bands appear fainter in reflectance, while the band crossing is at approximately the same position as in (b). (d-e) The same Fourier gratings as in (a-b), respectively, obtained before transferring to the glass substrate. The hBN flake is on a 90 nm $SiO_2$/Si substrate. On the $SiO_2$/Si substrate the waveguide resonances are associated with reflectance dips, and a modified background reflectance. The resonance linewidths are broader compared to the results obtained with the glass substrate. (f) Fourier grating with similar $A_1$ and $\Lambda$ as in (b) but larger $t$. The band crossing is at a lower energy compared to (b) and bands from a higher-order waveguide mode appear due to the increased thickness.

| Profile | $A_1$ (nm) | $A_2$ (nm) | $\Lambda$ (nm) | Peak-to-peak (nm) | Thickness (nm) | RMSE of fit (nm) | RMSE, unpatterned (nm) |
|---|---|---|---|---|---|---|---|
| **Single sinusoid** (Fig. 2 and Fig. S2b,e) | 8.5 (0.3) | — | 276.08 (0.08) | 17 | 123 | 1.95 (0.09) | 0.9 |
| **Double sinusoid** (Fig. 3, 4, and 5) | 33.7 (1.8) | 6.1 (1.4) | 294.9 (1.7) | 71 | 220 | 10 (1.8) | 9.8 |
| **Single sinusoid** (Fig. S2a,d) | 10.8 (0.6) | — | 319.6 (0.3) | 21.6 | 120 | 2.00 (0.18) | 0.9 |
| **Single sinusoid** (Fig. S2c) | 4.47 (0.23) | — | 276.29 (0.13) | 8.9 | 127 | 1.28 (0.09) | 0.9 |
| **Single sinusoid** (Fig. S2f) | 8.7 (0.6) | — | 275.6 (1.1) | 17.4 | 207 | 1.9 (0.4) | 1.0 |

**Table S1:** Fitted parameters of sinusoidal surface profiles from AFM topography measurements. Reported values are the mean (standard deviations) of single line fits to the grating regions. The thickness is the average thickness of the hBN flake or vdW heterostructure in the patterned region. The single sinusoid profiles are fitted to $f(x) = A_1 \cos(q_1 x)$, where $q_1 = 2\pi/\Lambda$. The double sinusoid profile is fitted to $f(x) = A_1 \cos(q_1 x) + A_1 \cos(q_2 x - \pi/2)$, where $q_1 = 2\pi/\Lambda$ and $q_2 = 2q_1$.

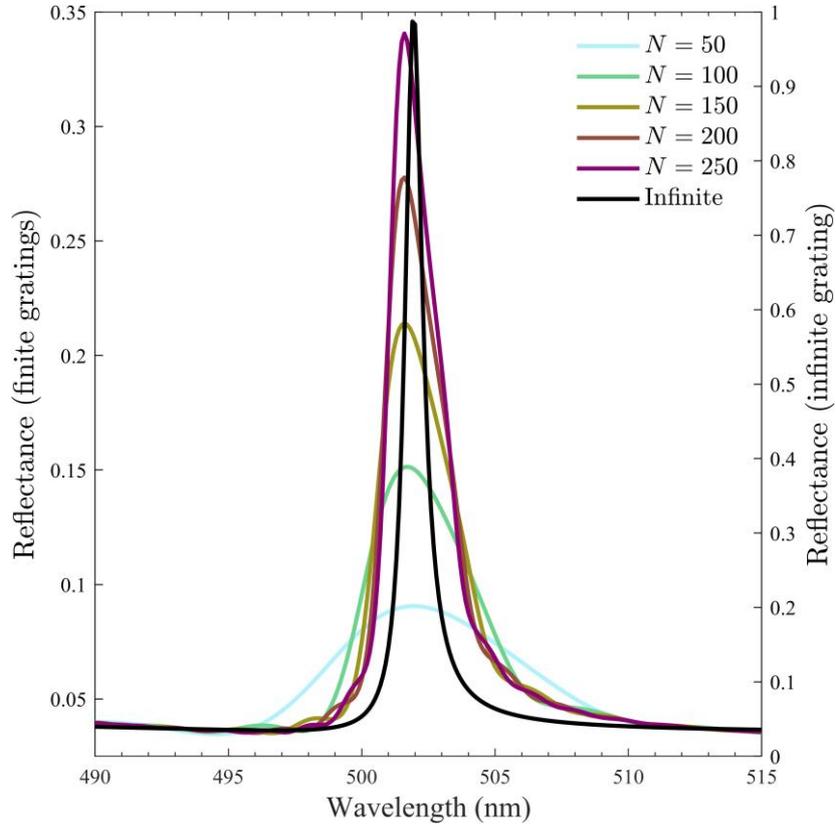

**Figure S3:** Simulated reflectance of single-sinusoid gratings with a finite number of periods *N* (left axis) and infinite periods (right axis). The simulated Fourier grating has the same amplitude, spatial period, and thickness as the one in Fig. 2 (see Table S1), and the incident and reflected light is at normal incidence ($k_x = 0$). The infinite case is realized with periodic boundary conditions. The finite case is realized with perfectly matched layers surrounding the simulation domain and a finite-sized port that launches a normally-incident plane wave. The wavelength of the reflectance maximum is nearly the same for the various grating sizes, while the lineshape broadens and the reflectance maximum drops when the number of periods decreases. We also observe small oscillations near the main reflectance peak, known as an Airy pattern, arising from the finite size of the grating, which acts as an aperture. The broadened lineshape and reduced reflectance maximum we observe in experiments compared to simulations with infinite gratings are therefore attributed to the limited experimental grating size (15x15 μm, corresponding to 54 periods).

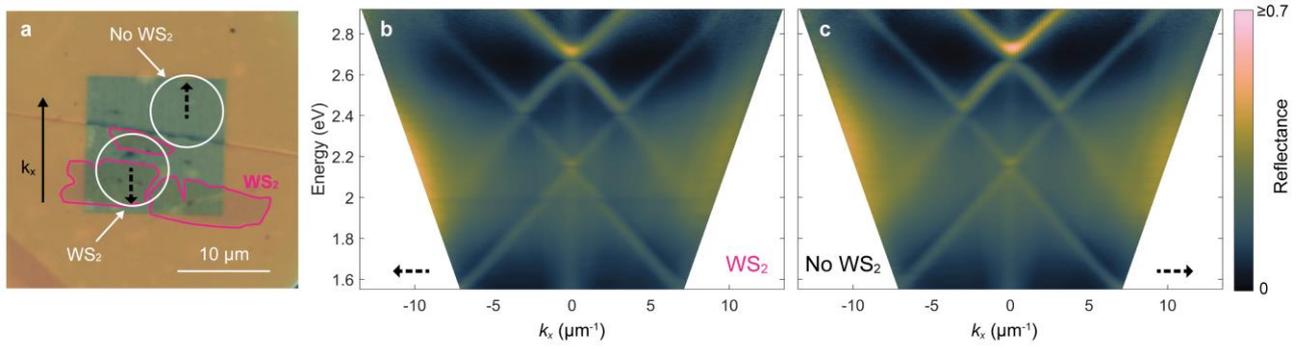

**Figure S4:** Propagation direction of waveguide modes with respect to the grating edge. (a) Bright field microscope image of the Fourier grating in the vdW heterostructure. Black solid arrow indicates the propagation direction of positive $k_x$. Black dashed arrows indicate the propagation direction that is towards the grating edge for measurements from the region with and without $WS_2$. (b) Angle-resolved reflectance from the region with $WS_2$ indicated by the white circle in (a). The reflectance is shown both for negative and positive $k_x$. Black dashed arrow indicates the propagation direction towards the grating edge. (c) Same as in (b) but for the grating region without $WS_2$. Both spectra show a slight asymmetry in reflectance intensity at positive and negative $k_x$ values, which is most clearly visible at higher energies, but it varies which side exhibits the largest intensity. This suggests that the small intensity differences are not a feature of the grating but arise from the different measurement positions. Modes that propagate towards the grating edge may reach the edge before coupling out, which would lead to slightly reduced reflectance.

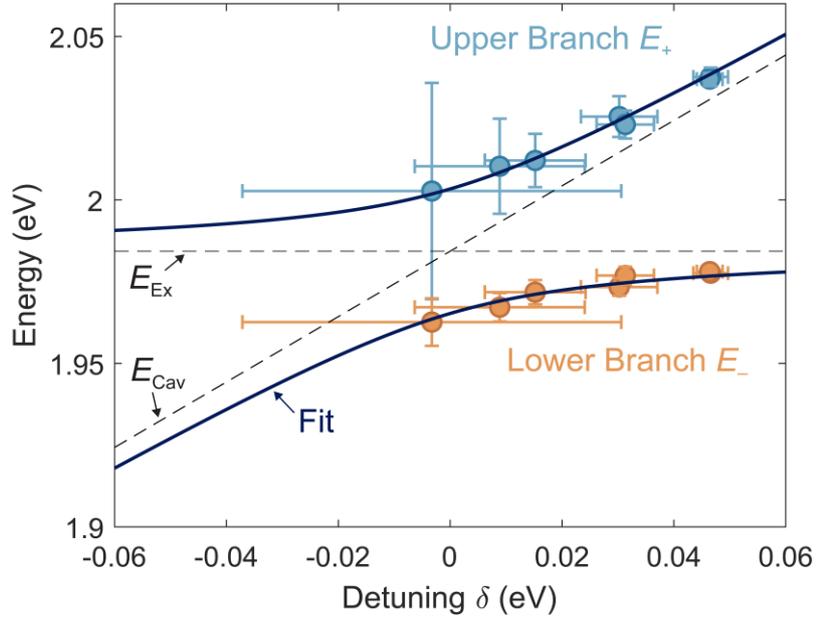

**Figure S5:** Detuning plot (energy vs. detuning $\delta = E_{Cav} - E_{Ex}$), where the uncoupled waveguide mode is estimated from energy conservation, $E_{Cav} = E_+ + E_- - E_{Ex}$. The upper (blue dots) and lower (orange dots) branches ($E_\pm$) are fitted to the coupled oscillator model (dark blue line) in the region where the branches overlap. The vertical error bars in the energy are given by the 95% confidence interval of the fitted values of $E_\pm$, while the horizontal error bars in detuning $\delta = E_+ + E_- - 2E_{Ex}$ are estimated from the propagation of errors. The dashed lines outline the uncoupled exciton ($E_{Ex}$) and waveguide modes ($E_{Cav}$).

|  | $E_{Rabi}$ | $\gamma_{Cav} + \gamma_{Ex}$ | $2g$ | $|\gamma_{Cav} - \gamma_{Ex}|$ |
| --- | --- | --- | --- | --- |
| $E_{Cav}$: hBN mode (meV) | 40 ± 9 | 48.9 ± 0.6 | 44 ± 8 | 18.7 ± 0.6 |
| $E_{Cav}$: Energy conservation (meV) | 38 ± 4 | 48.9 ± 0.6 | 42 ± 3 | 18.7 ± 0.6 |

**Table S2:** Fitted parameters of the uncoupled and coupled systems. Determined values of the $E_{Rabi}$ and $2g$ using two different approaches for approximating the energy of the uncoupled waveguide mode. The first approach assumes the uncoupled waveguide mode corresponds to the waveguide mode in only hBN (see Fig. 5). The second approach assumes the energy of the uncoupled waveguide mode can be determined by invoking energy conservation where the upper and lower branches overlap (see Fig. S5). A commonly used criterion to characterize the strong coupling regime is that $E_{Rabi} > \gamma_{Cav} + \gamma_{Ex}$ and $2g > |\gamma_{Cav} - \gamma_{Ex}|$, where $\gamma_{Cav(Ex)}$ are the half-width-at-half-maximum (HWHM) of the uncoupled hBN waveguide mode and exciton mode, respectively.

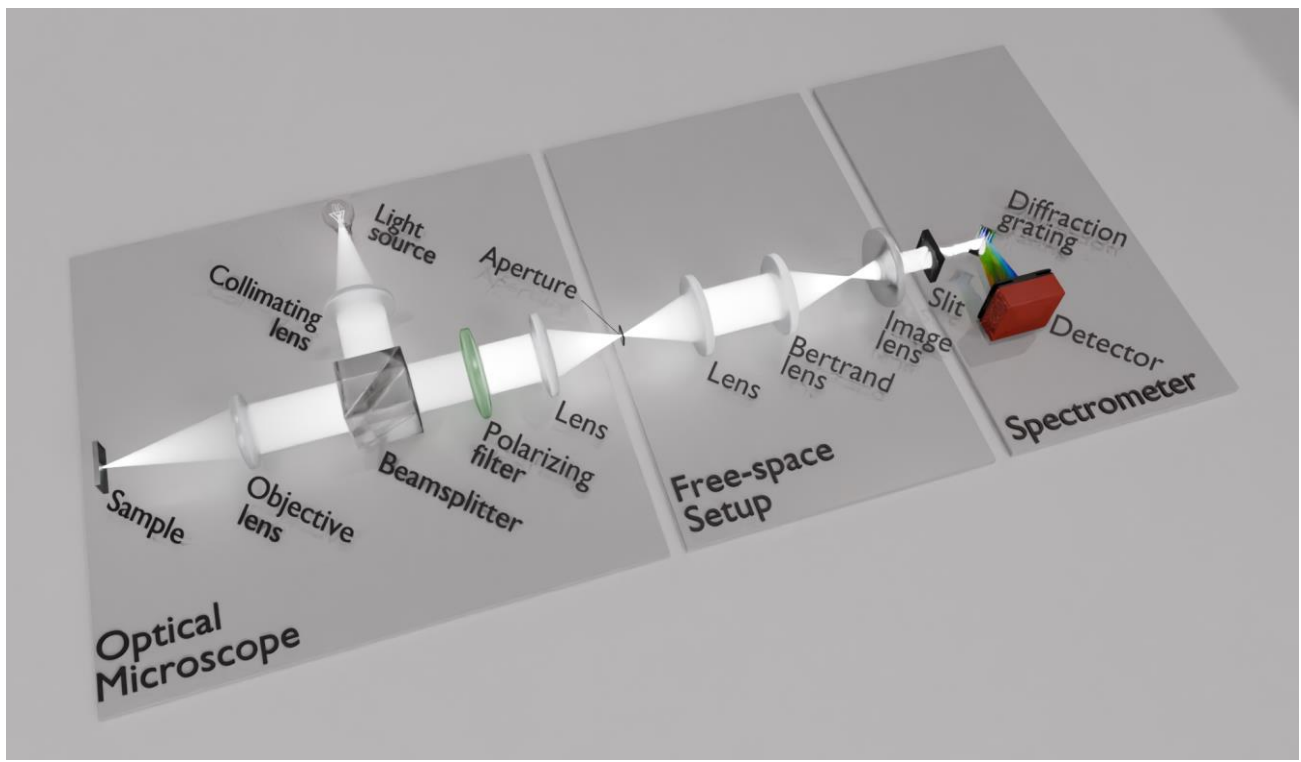

**Figure S6:** Schematic illustration of the optical setup used for reflectance measurements.

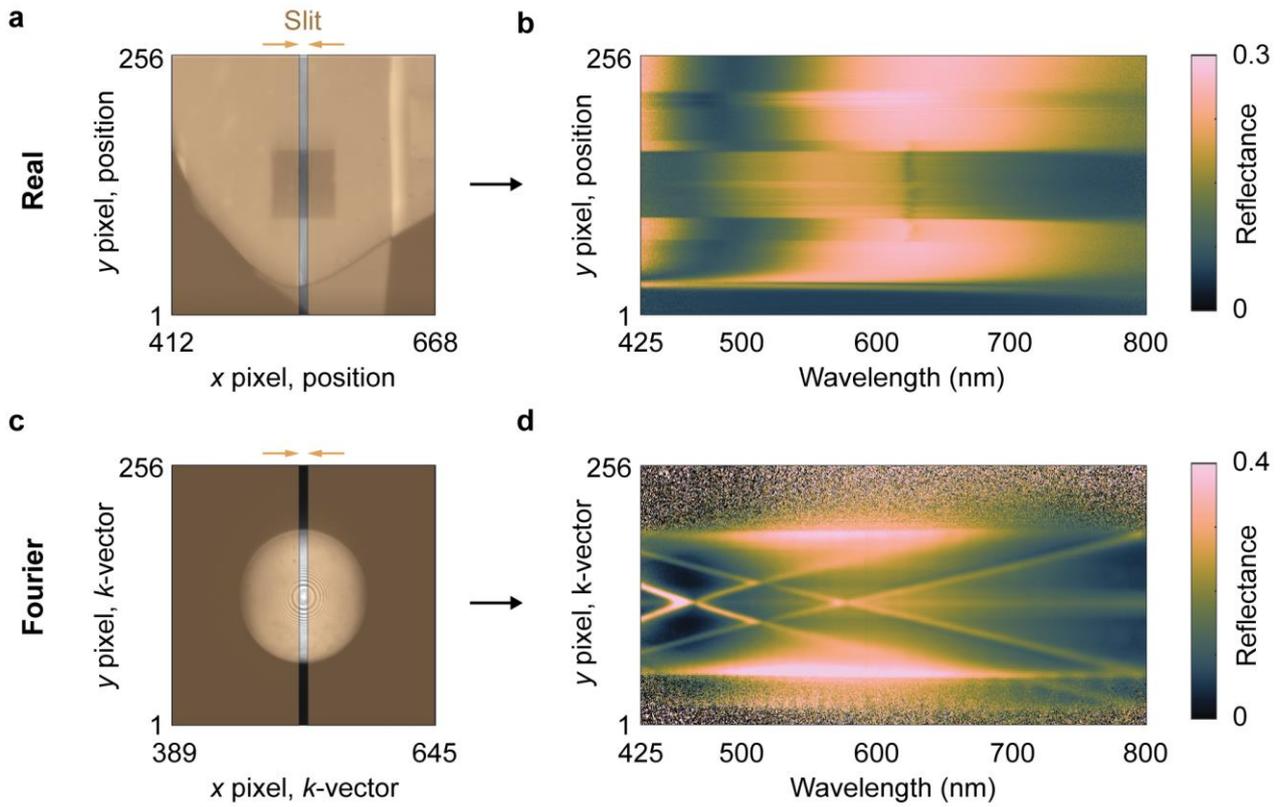

**Figure S7:** Real space and angle-resolved reflectance spectroscopy. (a) Real-space image of a sinusoidal grating in hBN, recorded by the CCD of the spectrometer. Shaded orange regions schematically illustrate the position of the slit. (b) Spectra obtained at each *y*-pixel along the slit in (a). The *y*-pixels correspond to a spatial position on the sample. (c) Fourier-transform of the real-space image of the grating obtained by inserting the Bertrand lens. The shaded orange regions again illustrate the position of the slit. (d) Angle-resolved reflectance spectra obtained at each *y*-pixel along the slit in (c). The *y*-pixels correspond to different *k*-vectors of the reflected light.